\documentclass[twoside]{LCWS11}
\usepackage[utf8]{inputenc}
\usepackage{graphicx}
\usepackage{wrapfig,rotating}
\usepackage{amssymb,amsmath,array}
\usepackage{cite}

\begin{document}
\title{
Forward Tracking in the ILD Detector} 
\author{\textbf{\textit{Robin Glattauer, Rudolf Fr\"uhwirth, Jakob Lettenbichler}} and \textbf{\textit{Winfried Mitaroff}}
\vspace{.3cm}\\
Austrian Academy of Sciences -- Institute of High Energy Physics\\
Nikolsdorfer Gasse 18, AT-1050 Vienna, Austria
}

\maketitle

\begin{abstract}
The reconstruction software for ILD is currently subject to a major revision, aiming at improving its accuracy, speed, efficiency and maintainability in time for the upcoming DBD Report. 
This requires replacing old code by novel methods for track search and fit, together with modern standards for interfaces and tools.

Track reconstruction in the ``forward region'', defined by the silicon Forward Tracking Detector (FTD), relies heavily on a powerful stand-alone track search. The new software makes use of a Cellular Automaton, a Kalman filter, and a Hopfield Neural Network. We give an overview of the project, its methods and merits.
\end{abstract}

\section{Introduction}

\begin{figure}[h]
  \centering
  \includegraphics[width=0.58\columnwidth]{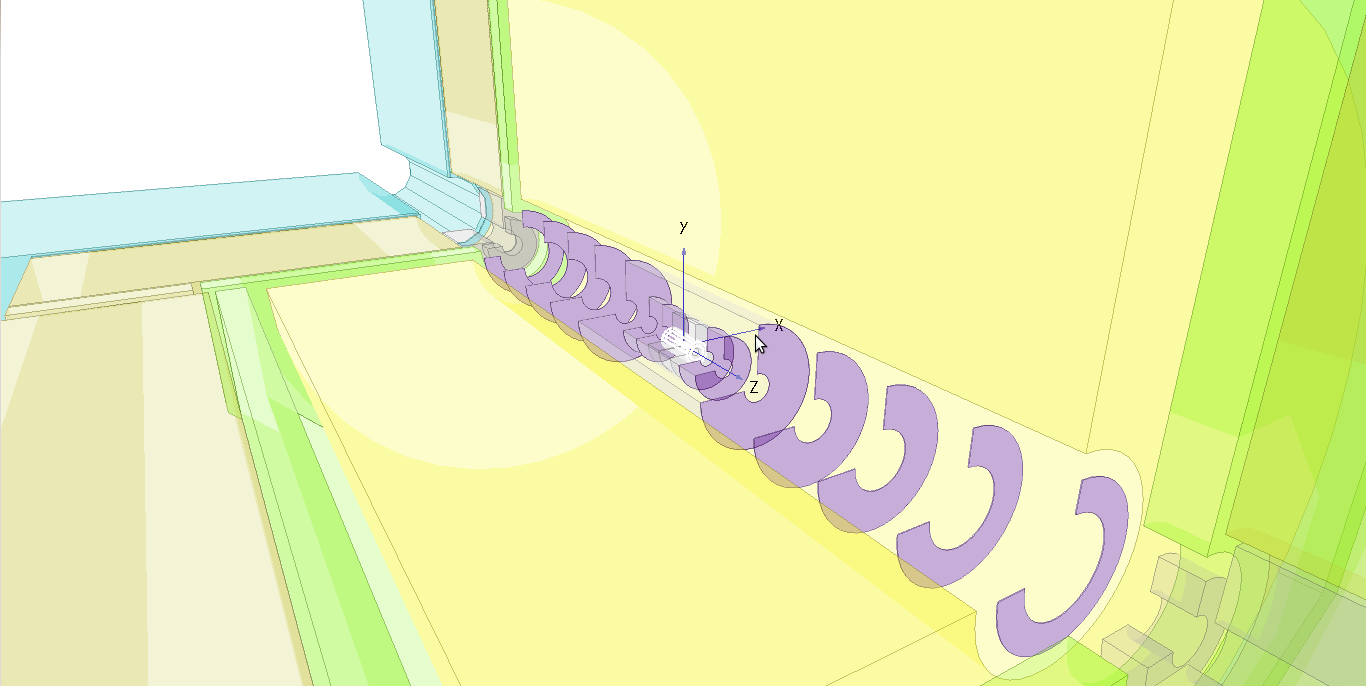}
  \includegraphics[width=0.41\columnwidth]{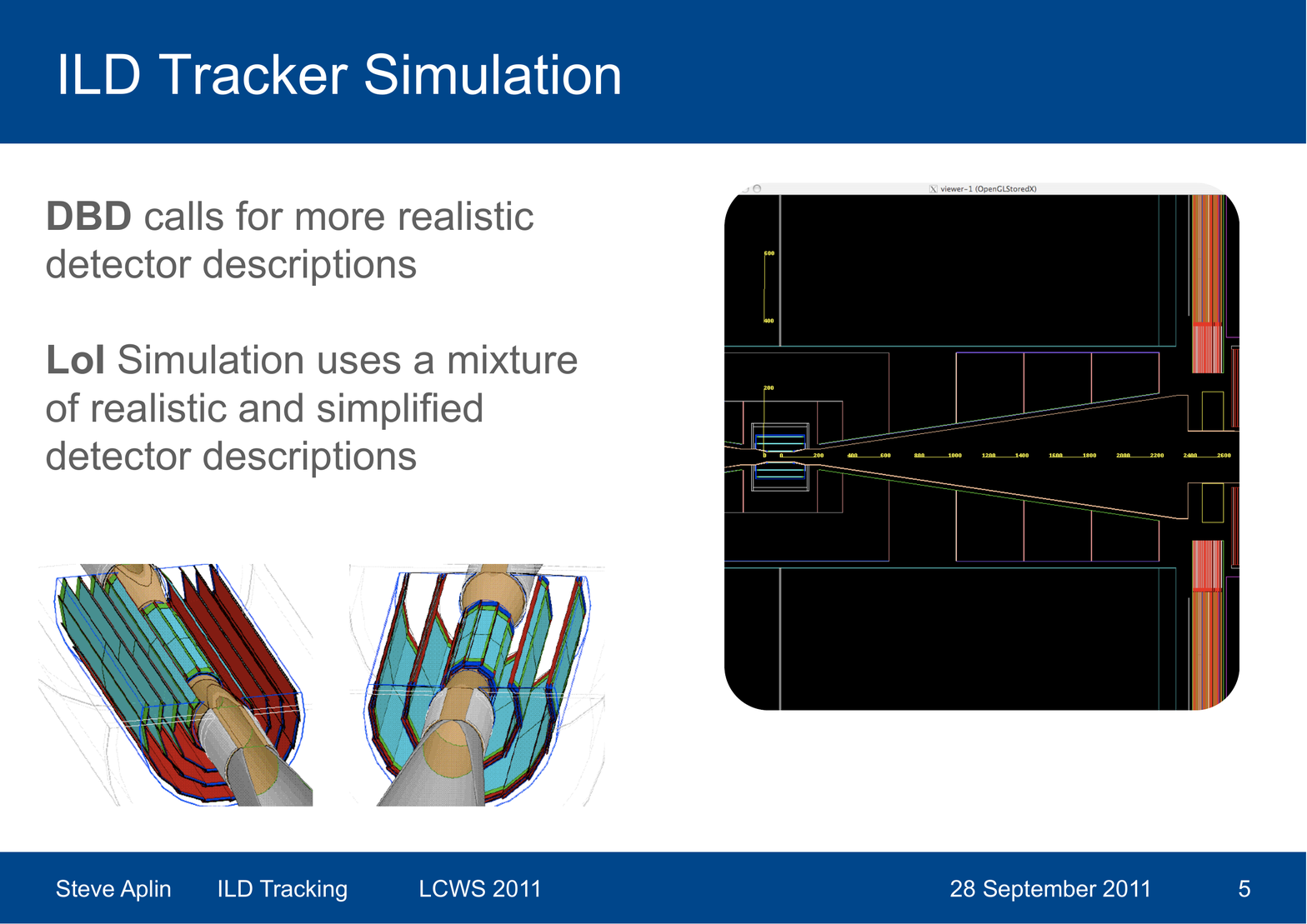}
  \caption{\textit{The FTD -- 3D view (left), and longitudinal section of one half (right).}}%
  \label{fig:FTD}%
\end{figure}

The Forward Tracking Detector (FTD) covers the region between the beam-cones and the central TPC of the International Large Detector (ILD) \cite{ILD_LoI}. Each half consists of 2 silicon pixel disks and 5 double-sided Si micro-strip disks, Fig.~\ref{fig:FTD}.
Tracks in that region are reconstructed within ILD's software framework \texttt{Marlin} \cite{iLCsoft}. 
Up to now this has been achieved by an older version of track search (\texttt{SiliconTracking}), together with a track fit of legacy Fortran code which has recently been replaced by the new Kalman Filter toolkit \texttt{KalTest} \cite{KalTest}. 
However, it is no longer adequate in terms of maintainability, flexibility and performance. 

Our goal is to develop a state-of-the-art standalone forward tracking package to process the hits on the FTD, in order to efficiently find and precisely reconstruct the genuine tracks. 
It is called \texttt{ForwardTracking}, and its methods are, in the order as they are used:

\begin {itemize}
  \item a \textbf{Cellular Automaton} for finding track candidates from the detected hits; 

  \item a \textbf{Kalman Filter} for fitting the track candidates and getting a quality feedback;

  \item a \textbf{Hopfield Neural Network} for getting the best subset of the track candidates collected  so far.

\end {itemize}

\section{The Cellular Automaton}            


In information theory, a Cellular Automaton is a set of discrete entities called ``cells'', which have discrete states 
and change those states -- in discrete iterations -- depending on their ``environment''.
E.g., biological cells are discrete because they always consist of an integer number, and are assigned discrete states like ``living'' or ``dead''. 
Maybe the most famous application of a Cellular Automaton is \textit{J.H.~Conway}'s ``Game of Life''.


Besides simulating cells or population growth, the Cellular Automaton can as well be used as a tool for pattern recognition \cite{CATS}.
While it is hard to define ``pattern'' exactly, what we might say is that a pattern follows some well-defined rules. 
A Cellular Automaton is based on rules, too: the rules on how to change the states of its cells. 
Thus, by implementing rules in such a way that they resemble the rules of a pattern being searched for, the Cellular Automaton can be used to find cells that form that specific pattern.

\subsection{A Toy Example of Track Search}

In track search, the pattern which is to be found is that of a trajectory belonging to a charged particle moving in a stationary magnetic field. In case of a homogeneous field and neglecting material effects (multiple scattering and energy loss), this trajectory is a helix with its axis parallel to the field \cite{Springer}.
Moreover, our task is simplified by the geometry of the FTD being approximately planes perpendicular to the helix axis.

\begin{figure}[htbp]%
  \begin{minipage}{0.5\columnwidth}%
  \includegraphics[width=1\columnwidth]{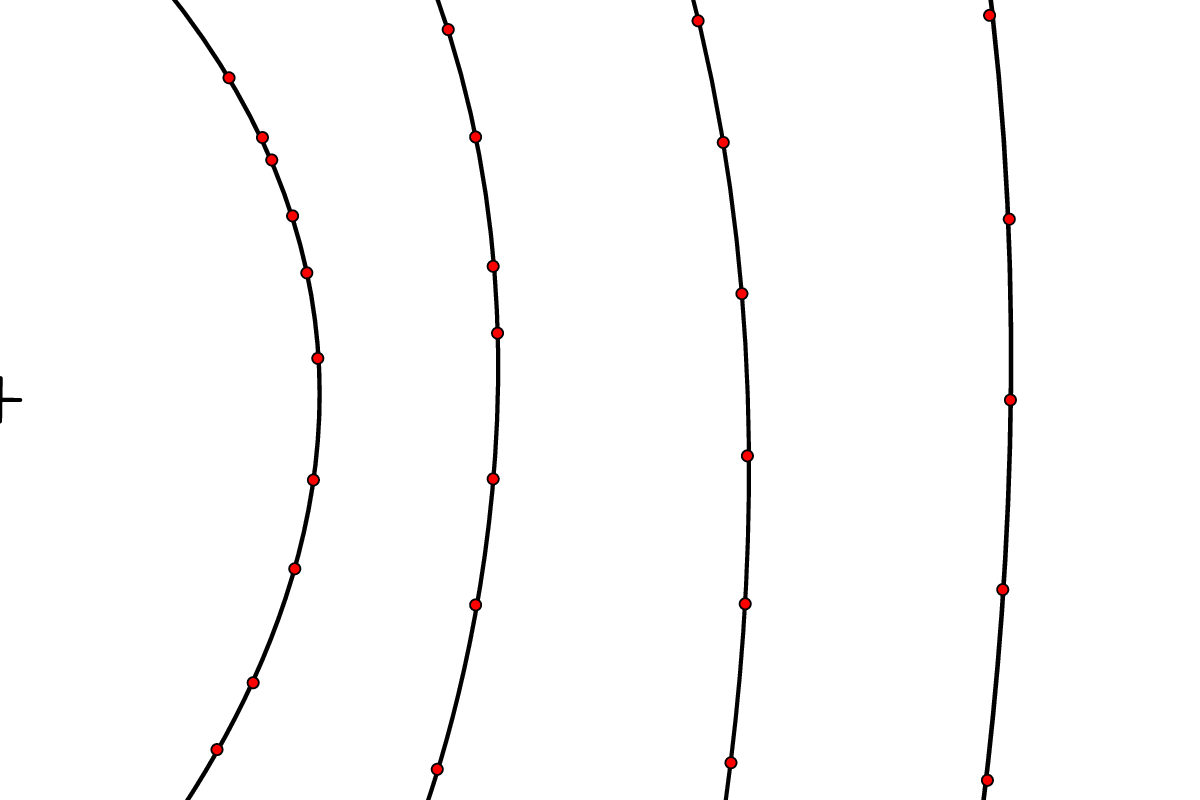}
  \caption{\textit{The hits.}}%
  \label{fig:detector}%
  \end{minipage}%
  \begin{minipage}{0.5\columnwidth}%
  \includegraphics[width=1\columnwidth]{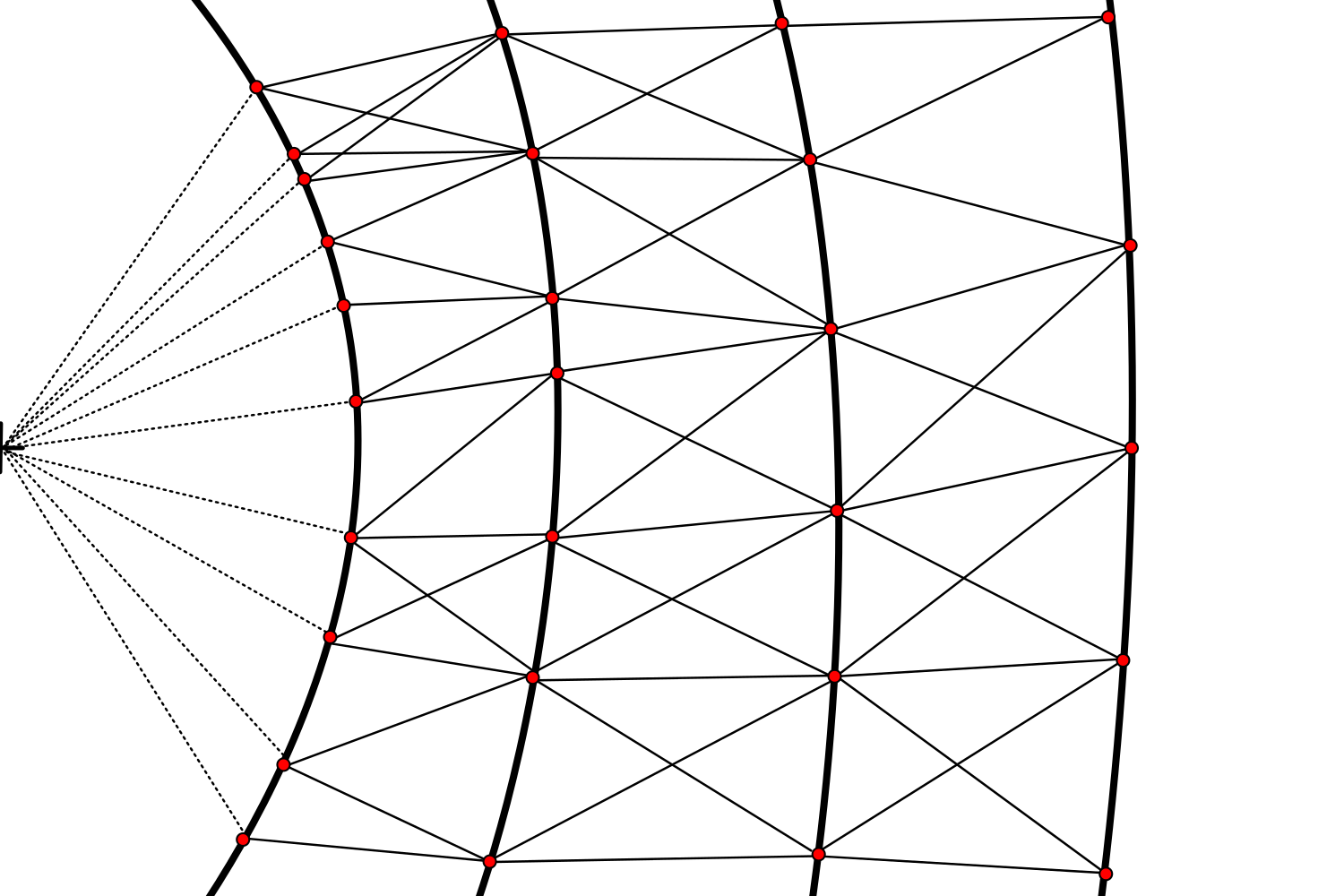}
  \caption{\textit{Segments built.}}%
  \label{fig:RPhi_step1}%
  \end{minipage}%
\end{figure}

In order to explain the method of the Cellular Automaton, for a clearer picture let's reduce the problem to 2D, and imagine a hypothetical detector of 4 layers as in Fig.~\ref{fig:detector}. 
The half-cross on the left indicates the centre of the beam interaction profile which is used as an additional ``hit'' for track search; this is valid for most tracks, since they originate from near that point -- either directly, or as the decay of a short-lived particle. 

At the core of any Cellular Automaton are the cells. In track search, the cells are defined as the connection of two (or more, see later) hits between adjacent layers, usually called a ``segment''.
The first step is to build all the segments we are going to work with. 
However, connecting all the hits between adjacent layers, i.e.~building every possible segment, is unnecessary and combinatorially expensive.
Fortunately we have some idea, supported by simulation, how the tracks we are looking for are expected to behave. 

Our strategy is to throw away as much fake data as possible with fast test criteria early; the lesser data we retain, the more elaborate but time consuming tests can we apply later. Of course we have to make sure that those early fast test criteria are justified by a careful analysis of representative samples of realistically simulated tracks.

For this example we apply as a very simple criterion the distance between two hits: if too far apart they will not get connected.\footnote{
This may sound rather blunt. But it is a valid criterion, and it is very fast.}
After connecting only those hits between adjacent layers whose distance is less than a given cut-off value, we arrive at Fig.~\ref{fig:RPhi_step1}: 
the lines connecting the hits are the segments we will further work with.

The cells of a Cellular Automaton have states. For track search, the state is expressed as an integer value. 
At start-up, all segments are given an initial state value 0 (indicated by black in Fig.~\ref{fig:RPhi_step1}).
The Cellular Automaton will change those states by the following rule:
in each iteration, it checks all segments for ``neighbours'' -- if a segment has at least one neighbour, its state gets augmented by 1; otherwise, its state stays the same.

A ``neighbour'' is defined as an \underline{inward} segment if
\begin{itemize}
 \vspace{-0.5ex}
 \setlength{\itemsep}{-1ex}
 \item it shares a hit with the current segment, and
 \item has the same state as the current segment, and
 \item fulfills some test criteria (representative for the track we are looking for).
\end{itemize}
                                                                                      
Criterion in this example is the angle between the segment and its candidate neighbour: 
a small angle (close to a straight line $\Rightarrow$ big helix radius, high $p_T$) is ``pass'', 
whereas a big angle (a pronounced kink $\Rightarrow$ small helix radius, low $p_T$) is ``fail''.\footnote{
This is good only for stiff tracks, and is of limited power; it serves rather for illustration. There are more powerful criteria, also ones that are more suited for low-momentum tracks.} 

\begin{wrapfigure}{r}{0.5\columnwidth}
\centerline{\includegraphics[width=0.4\columnwidth]{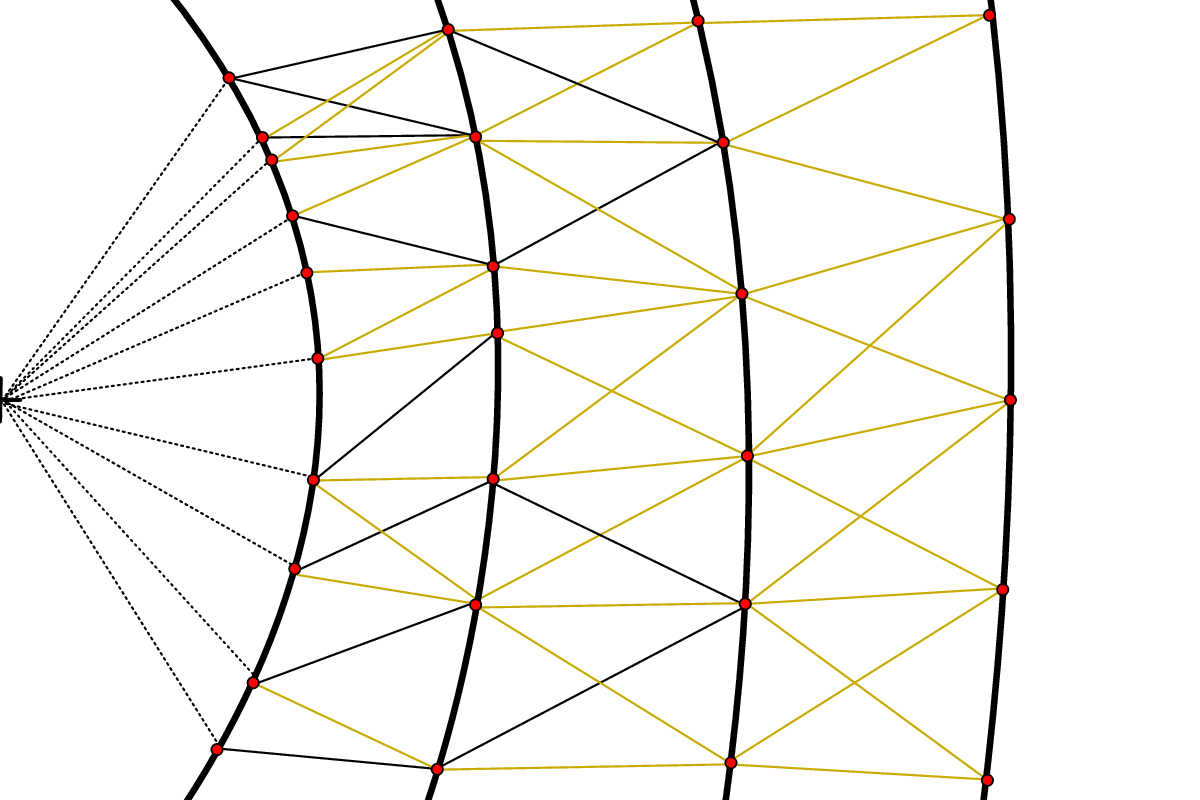}}
\caption{\textit{After first iteration.}}\label{fig:RPhi_step2}
\end{wrapfigure}


After the first iteration we get Fig.~\ref{fig:RPhi_step2}: 
the segments in yellow are those which have  been raised to state 1.
Note that the innermost segments (the ones connected to the beam interaction point) all stay at state 0, because they don't have inward segments sharing their pseudo hit, thus no neighbours have been possible for them. 

\begin{figure}[htbp]%
  \begin{minipage}{0.5\columnwidth}%
  \includegraphics[width=1\columnwidth]{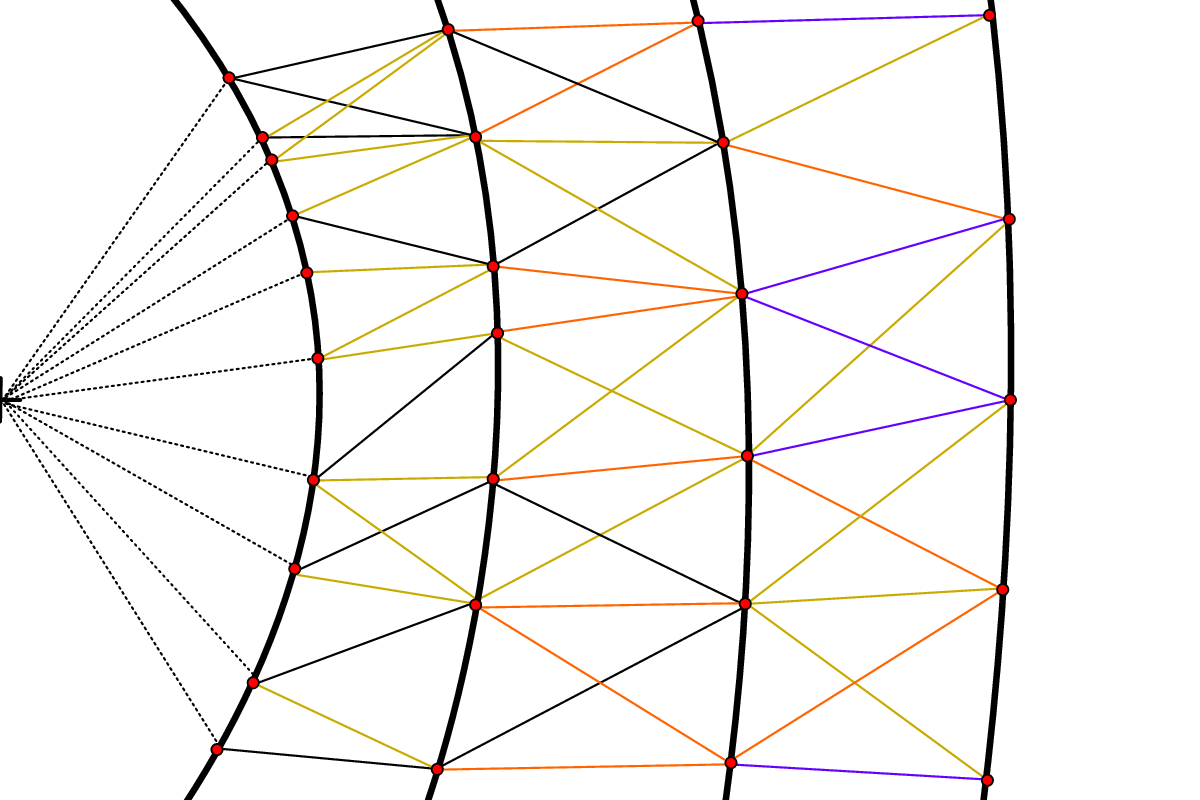}
  \caption{\textit{End of the iterations.}}%
  \label{fig:RPhi_step2_3}%
  \end{minipage}%
  \begin{minipage}{0.5\columnwidth}%
  \includegraphics[width=1\columnwidth]{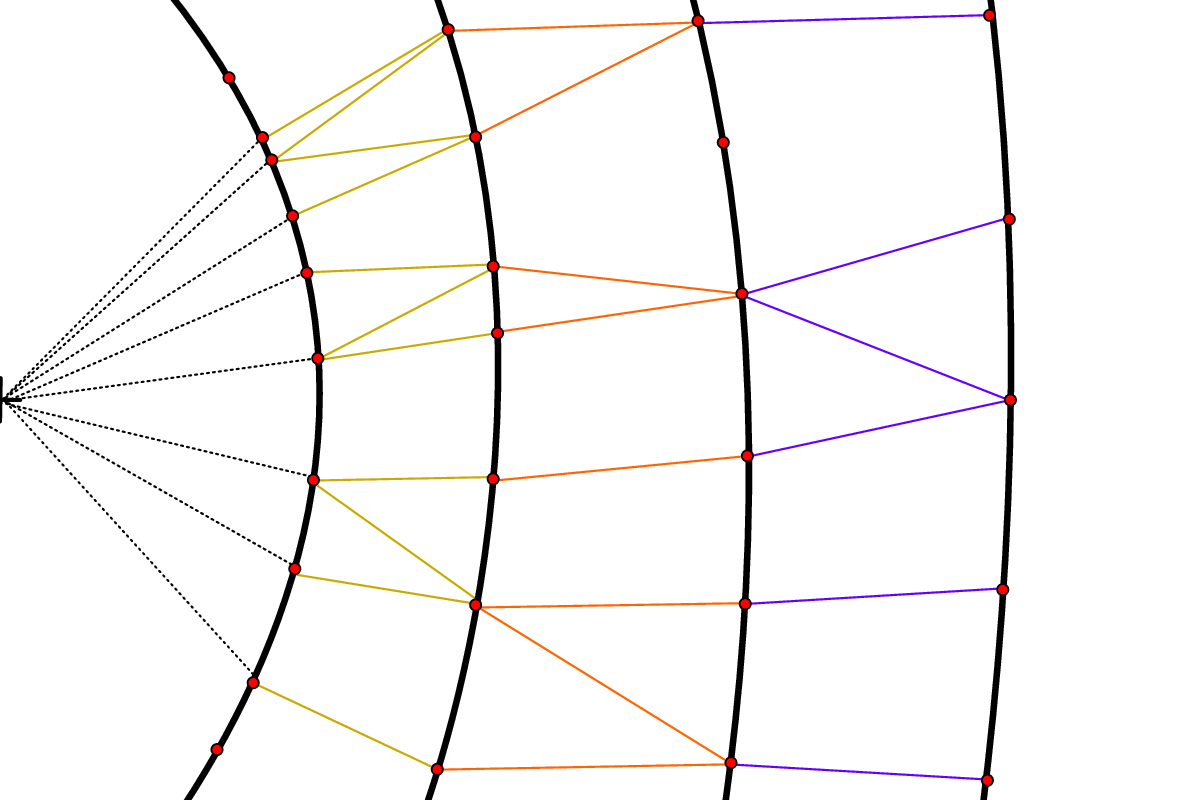}
  \caption{\textit{Bad segments erased.}}%
  \label{fig:RPhi_step3}%
  \end{minipage}%
\end{figure}

The Cellular Automaton continues iterating until no changes happen anymore. Eventually we arrive at Fig.~\ref{fig:RPhi_step2_3}: there are segments of state 0 (black), state 1 (yellow), state 2 (red), and state 3 (purple). 

Let's label our $n = 4$ detector layers from $i = 1$ (the innermost) to $i = n$ (the outermost), and call a segment to be ``from layer $i$\,'' if it connects hits between the layers $i$ and $i + 1$.
Now, if we take any outermost segment (i.e.~one from layer 3) of state 3, it is guaranteed to be linked in a chain to a segment of state 2 from layer 2, to a segment of state 1 from layer 1, and to a segment of state 0 from the interaction point. Thus, it is linked all the way through, which is the behaviour expected for the path of a genuine track.

On the other hand, any outermost segment (i.e.~one from layer 3) which doesn't have state 3 is \underline{not} linked all the way through, and consequently gets thrown away. 
In general, all segments which have not attained their highest possible state -- which is exactly identical to the label of their ``from'' layer as defined above -- will be erased.


After the clean-up we finally arrive at Fig.~\ref{fig:RPhi_step3}: quite a number of possible segments have been erased, and we are left with few candidate paths. 
In a realistic scenario, particularly with background hits, such data reduction would become even much more pronounced; 
this notwithstanding, the number of candidate paths might still be too high. 

Therefore, the Cellular Automaton method shown so far can be further improved by augmenting the 2-hit segments to longer 3-hit segments. This allows for refined test criteria of stronger power, albeit also demands increased processing time.

\subsection{The Cellular Automaton for the FTD}

\begin{figure}[h]%
  \begin{minipage}{0.3\columnwidth}%
  \includegraphics[width=1\columnwidth]{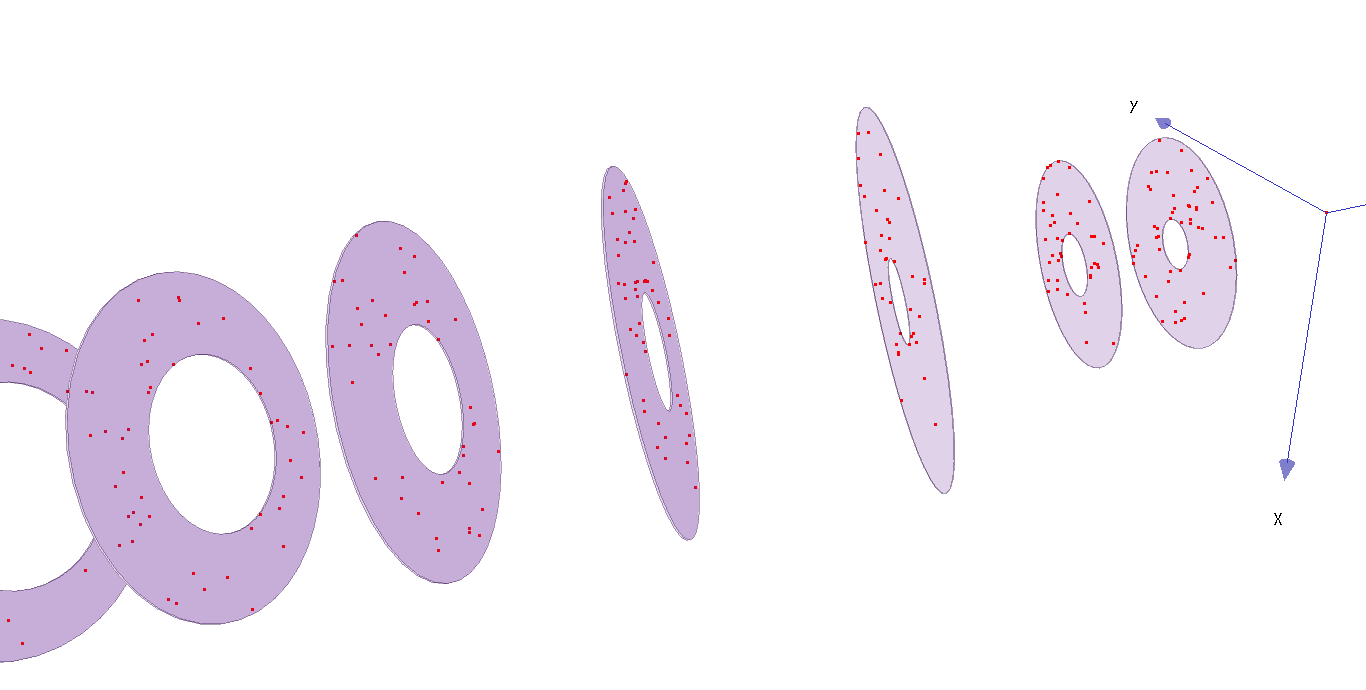}
  \caption{\textit{The hits.}}%
  \label{fig:1figsA}%
  \end{minipage}%
  \qquad
  \begin{minipage}{0.3\columnwidth}%
  \includegraphics[width=1\columnwidth]{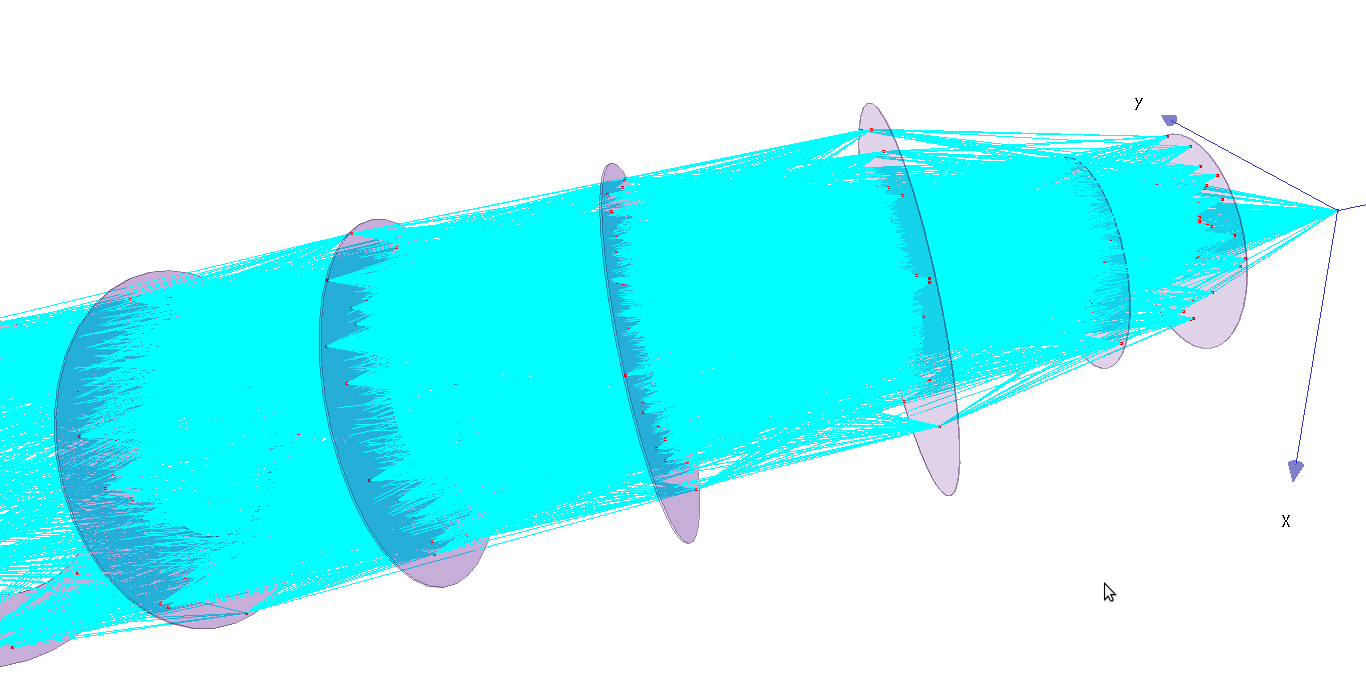}
  \caption{\textit{Segments built.}}%
  \label{fig:1figsB}%
  \end{minipage}%
  \qquad
  \begin{minipage}{0.3\columnwidth}%
  \includegraphics[width=1\columnwidth]{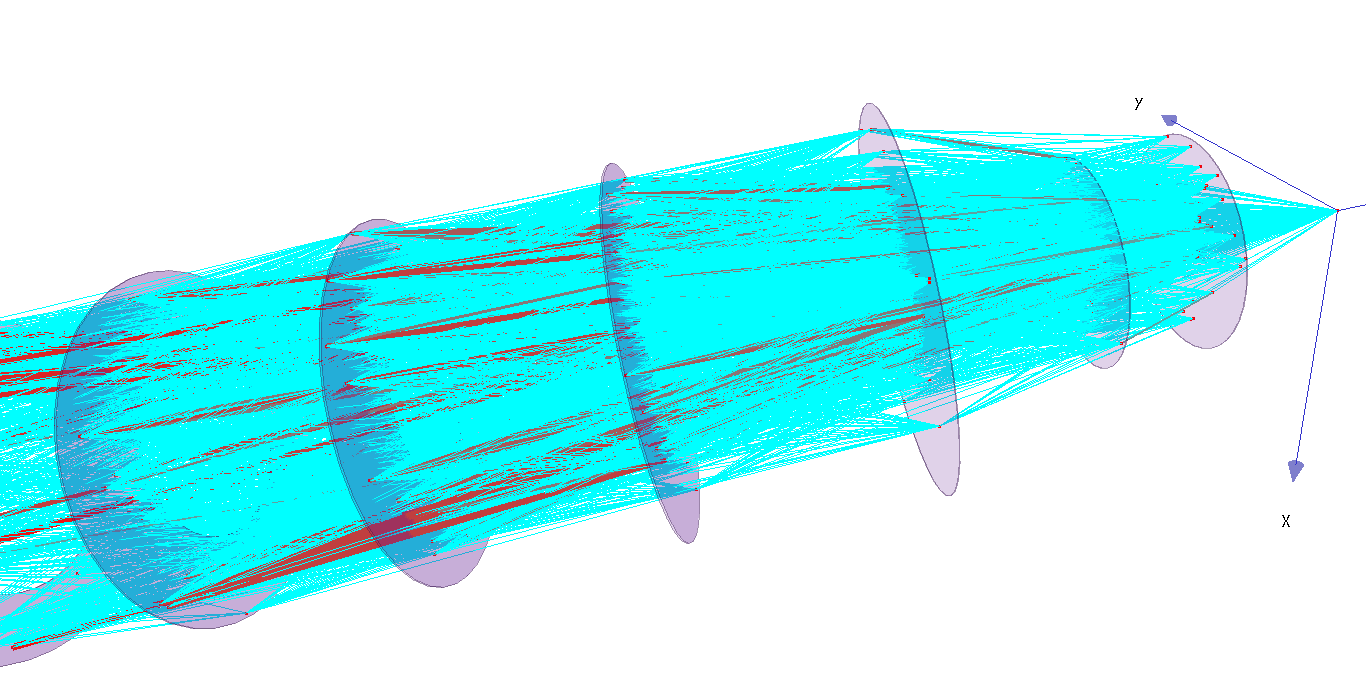}
  \caption{\textit{C.A. performed.}}%
  \label{fig:1figsC}%
  \end{minipage}%
\end{figure}

\begin{figure}[h]%
  \begin{minipage}{0.3\columnwidth}%
  \includegraphics[width=1\columnwidth]{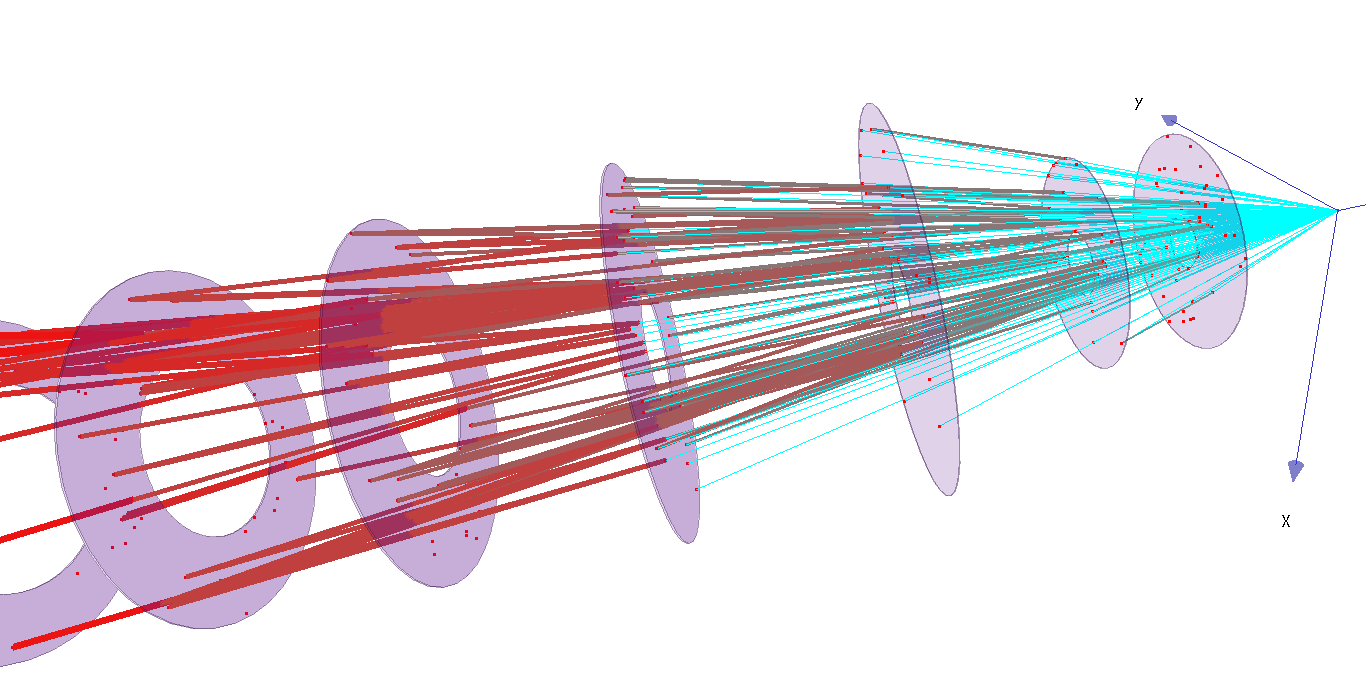}
  \caption{\textit{Bad segments erased.}}%
  \label{fig:2figsA}%
  \end{minipage}%
  \qquad
  \begin{minipage}{0.3\columnwidth}%
  \includegraphics[width=1\columnwidth]{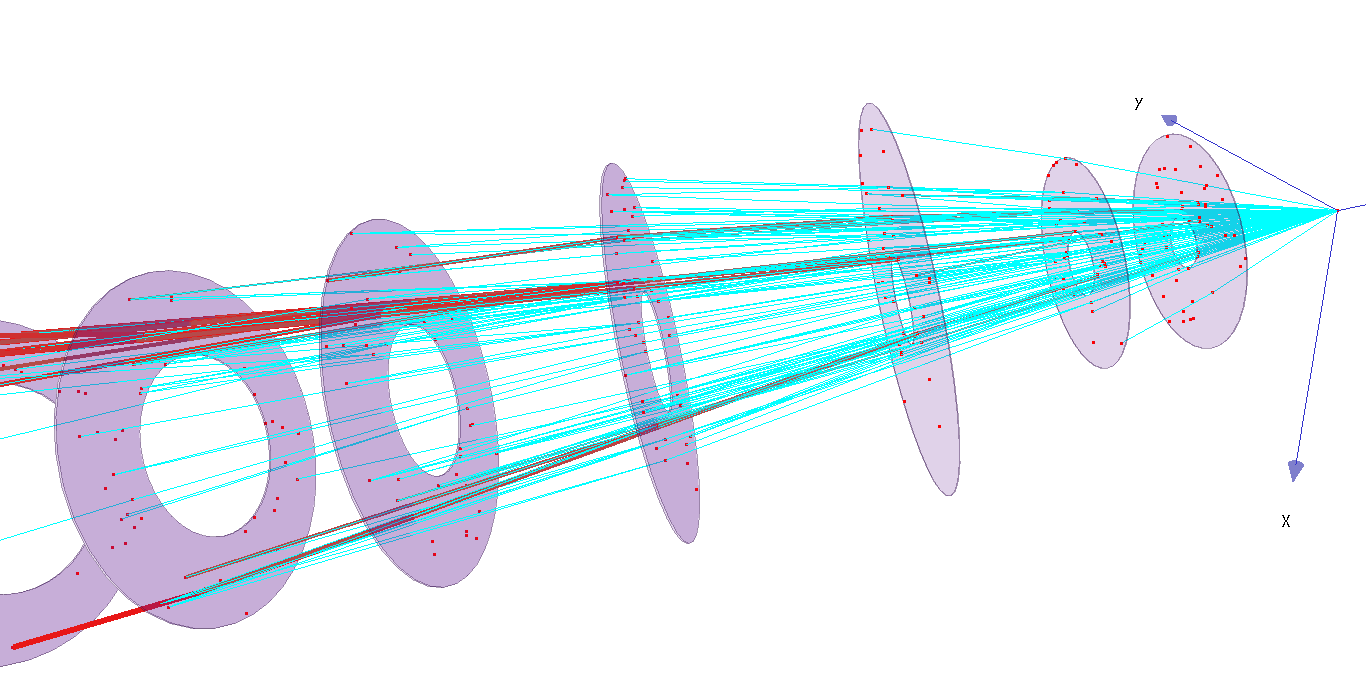}
  \caption{\textit{C.A. performed with longer segments.}}%
  \label{fig:2figsB}%
  \end{minipage}%
  \qquad
  \begin{minipage}{0.3\columnwidth}%
  \includegraphics[width=1\columnwidth]{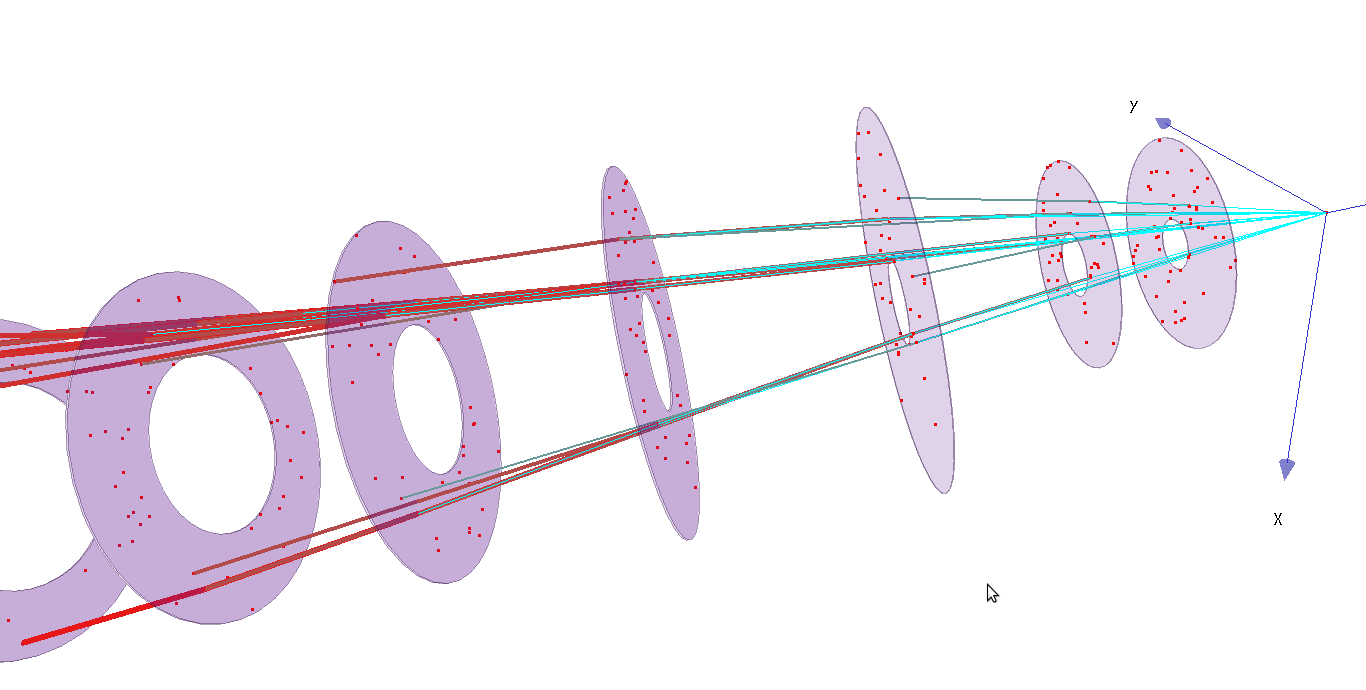}
  \caption{\textit{Bad longer segments erased.}}%
  \label{fig:2figsC}%
  \end{minipage}%
\end{figure}

The methods sketched above are implemented in the \texttt{ForwardTracking} package. 
Results for realistic physics events in the full FTD setup are shown in Figs. \ref{fig:1figsA} -- \ref{fig:2figsC}. 
Here, additional to the genuine hits, random salt-and-pepper hits have been distributed over the disks in order to also simulate some background. 


The Cellular Automaton is performed twice: first with 2-hit segments, and thereafter with longer 3-hit segments and refined test criteria. 
The states of the segments are indicated by color and thickness: the more reddish and thicker a segment is, the higher is its state. 
For details, refer to \textit{R.~Glattauer}'s talk at this conference.

\section{The Kalman Filter}

The Kalman Filter is a well-known and widely used technique for track fitting \cite{Rudi_Kal, Strandlie, Springer}. 
Estimating the track parameters is useful at a late stage of track search, as it gives knowledge about the quality of the track.\footnote{
An ultimate track fit by a Kalman Filter with smoother will also be performed, after track search, on the final sample, yielding the best estimate of track parameters everywhere along the trajectory.}
A common test criterion is the upper-tail $\chi^2$ probability of the fit: 
if it is very low, the track is most likely not a genuine one. 

However, a bit of caution is needed, because tracks might not behave as they ideally should. 
E.g., the mistaken addition of a wrong hit that made sense, or a charged pion decay that generates a small kink, may pass the Cellular Automaton stage, but will get excluded for its bad $\chi^2$ probability. 
This is known as  the ``outlier problem''.

The candidate paths left over by the Cellular Automaton are subjected to the Kalman Filter, and tested against a 
given cut-off value for their $\chi^2$ probability.

Technically, the Kalman Filter is called via the interface \texttt{MarlinTrk} of the framework \cite{iLCsoft}, providing the fitter toolkit: 
\texttt{KalTest} (core package doing the actual fitting) and \texttt{KalDet} (implementing detector geometry and materials) \cite{KalTest}. Thus, their possible replacement by some other fitter toolkit would be transparent to \texttt{ForwardTracking}.

\section{The Hopfield Neural Network}

\begin{figure}[htbp]%
 \begin{minipage}{0.5\columnwidth}%
 \includegraphics[width=1\columnwidth]{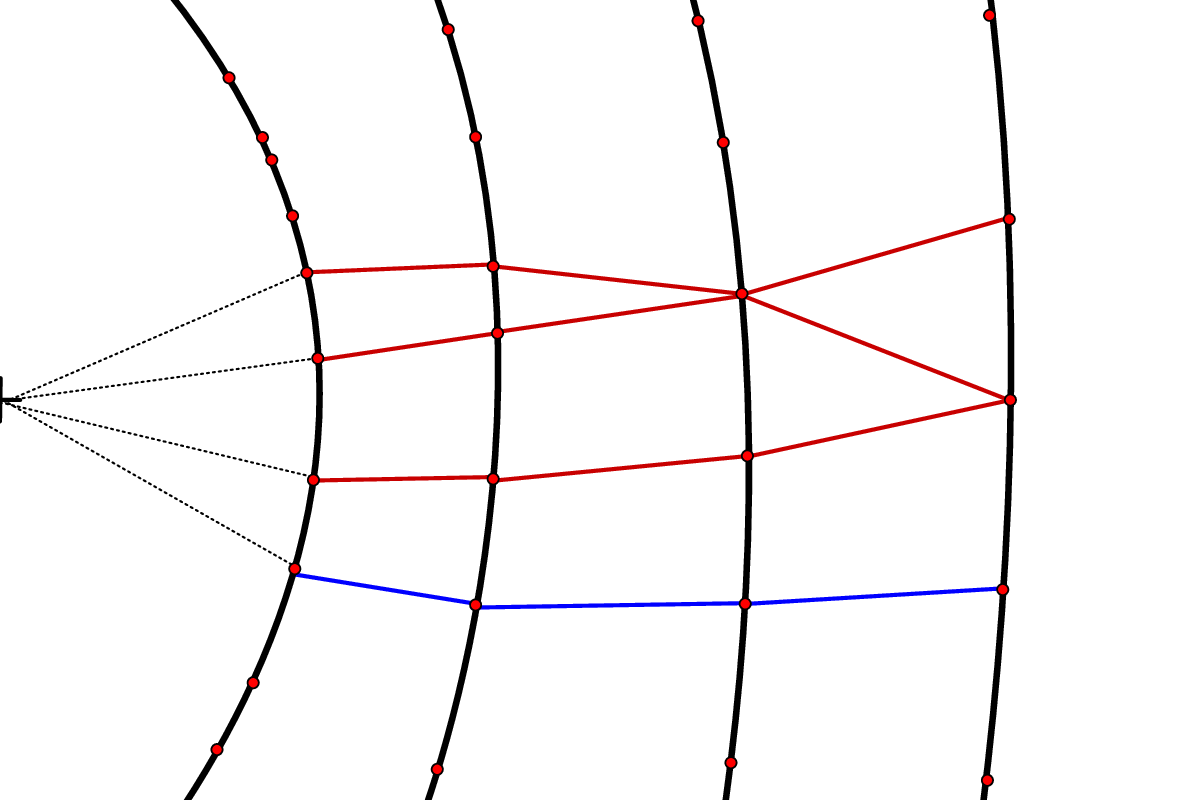}
 \caption{\textit{Conflicting tracks.}}%
 \label{Fig:RPhi_tracks_in_conflict}%
 \end{minipage}%
 \begin{minipage}{0.5\columnwidth}%
 \includegraphics[width=1\columnwidth]{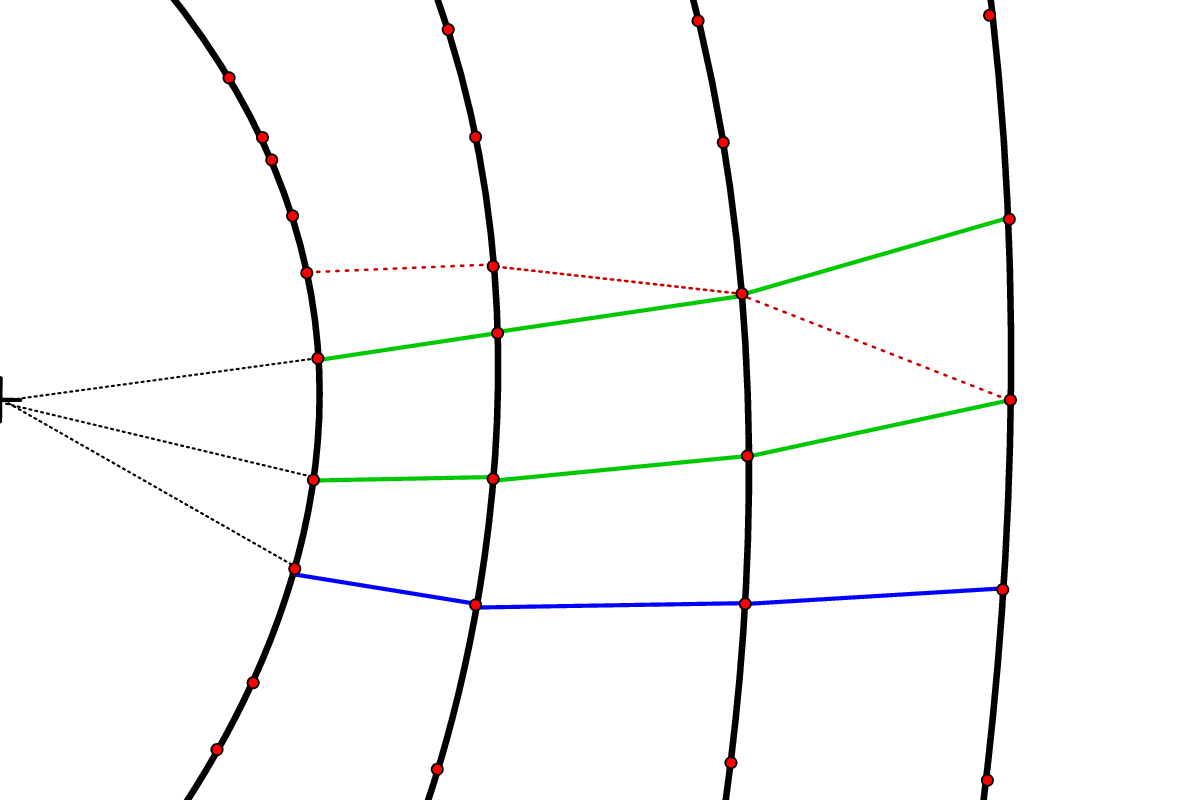}
 \caption{\textit{The best subset.}}%
 \label{Fig:RPhi_step6}%
 \end{minipage}%
\end{figure}

After cuts based on the Kalman Filter $\chi^2$ probability we have gathered track candidates with sufficient quality. 
However, before saving them, we have to make sure that those tracks are compatible, i.e.~do not share hits. 
Looking again at our 2D toy detector, we find 4 tracks remaining  (Fig. \ref{Fig:RPhi_tracks_in_conflict}).

The blue track does not share hits with any other, and can therefore be finally saved. 
But the three red ones share hits -- they are in conflict; thus we have to take a decision about which ones to keep and which to discard. 

A straightforward approach would keep the track with the best $\chi ^2$ probability (or another useful quality
indicator) and discard all tracks in conflict with it.
A situation might arise, however, where several tracks are entangled in a complicated way, and choosing the best quality indicator leads to the loss of other good tracks. 
Our aim is to find a subset that contains as many compatible tracks as possible, and has a sum of quality indicators as high as possible as well. 
As shown in~\cite{Rudi_NN} one can come close to such an ``optimal subset'' by using a so-called Hopfield Neural Network (HNN)~\cite{Shriv}.

The HNN is a dynamic system, composed of ``neurons'' which interact with each other. In our specific application
the neurons represent the tracks. Two tracks (neurons) can be either compatible or incompatible, i.e.~don't share hits or do so, respectively. The state of a neuron is a number between 0 and 1, where a higher value represents a larger probability of the neuron to be included in the final selection. 

The evolution of the network proceeds in discrete iterations. In every iteration the state of each neuron is recomputed by an ``activation function'' depending on the states of the other neurons. 
The connections (weights) between the neurons are set up in such a way that compatible neurons reinforce each other, and incompatible ones suppress each other. 
The quality indicator of the track is used in the activation threshold, such that neurons corresponding to higher quality tracks are activated more easily.

There are two problems that may arise in the evolution of such a dynamic system. First, it could be stalled in an infinite loop: the states of all neurons oscillating back and forth between two configurations. 
This problem can be avoided by asynchronous updating, i.e.~updating the neurons one at a time and preferably in random order.
The second problem is the presence of local minima, i.e.~convergence to a suboptimal configuration. 
This can be mitigated, if not totally avoided, by ``annealing'': 
the activation function now depends on a temperature parameter which is cooled down in every iteration. 

The network starts up with all neurons having as their state a small random value (e.g.~between 0 and 0.1).
In every iteration the states change their values; the network is considered to be stable as soon as these changes are sufficiently small (below a user defined cut-off value). 
Once this is attained, neurons with a state value above a certain threshold (e.g.~0.75) are considered as active and enter the best subset. 

The performance of the HNN depends on several parameters which have to be tuned for the concrete application in order to give the best results. 
In our example this results in two of the conflicting red tracks surviving and one being deleted (Fig.~\ref{Fig:RPhi_step6}).

\section{Expected Noise}


\subsection{Beam-induced Background}

Close to the beam interaction point electron-positron pairs are copiously produced by beamstrahlung, prominently in the forward direction. Those which escape the beam-cones affect the innermost detectors, i.e.~vertex detector and inner FTD disks. 
A rough estimate \cite{ILD_LoI} gives 10.4 (6.1) background hits per bunch crossing on disk 1 (2). 
These are Si pixel detectors \cite{Springer} with their readout integrating over many bunch crossings at 369 ns. 
A first guess about the final specifications expects a readout frame of 100 bunch crossings, resulting in the order of $10^3$~background hits per event on each of the inner FTD disks 1 and 2. 

\subsection{Artefact Ghost Hits}

The outer FTD disks 3 -- 7 are double-sided Si micro-strip detectors \cite{Springer}. The strips on either side are oriented to form a stereo angle; this allows for 2D measurements, but also introduces artefact ghost hits. 
For a stereo angle of $90^\circ$, $n$ genuine hits will cause up to $n(n-1)$ ghosts.
In this scenario, a dense jet of e.g.~10 tracks yields 90 ghosts on each disk, and on 3 disks yields $10^6$ triplets, only $10^3$ of which are not caused by ghosts; this would be a combinatorial disaster for the Cellular Automaton and must be avoided. 

For this reason the present design of the micro-strip disks uses a shallow stereo angle,\footnote{
Other effects of a shallow stereo angle (at radially oriented strips) are a better transverse momentum and impact parameter resolution, albeit worse longitudinal resolutions \cite{MV2008}.}
thus effectively ``pushing out'' most of the artefact ghost hits beyond the sensitive area of the disks and so avoiding to be included in the track search.

\section{Summary and Outlook}

A new software processor (\texttt{ForwardTracking}) has been developed for a stand-alone track search in the forward region of ILD,  covered by the silicon Forward Tracking Detector FTD. Methods used are a Cellular Automaton, a Kalman filter, and a Hopfield Neural Network. It is part of a major revision of ILD's software framework \texttt{Marlin}, and has successfully been tested and shown superior performance w.r.t.~the old software.

Future developments are planned to include a TPC supported track search in the ``intermediate region'' between barrel and forward, efficient outlier handling by a Deterministic Annealing Filter (DAF) \cite{Strandlie}, and a precision track fit for electrons -- which suffer energy loss by bremsstrahlung -- by a Gaussian Sum Filter (GSF) \cite{Adam, Strandlie}.

\section*{Acknowledgments}

The authors wish to thank \textit{Steven Aplin, Jan Engels} and \textit{Frank Gaede} (DESY, Hamburg) for their support with the integration into the ILD framework.


\begin{footnotesize}


\end{footnotesize}


\end{document}